\documentclass[conference]{IEEEtran}
\IEEEoverridecommandlockouts
\usepackage[colorlinks=true, linkcolor=blue, citecolor=blue, urlcolor=blue]{hyperref}
\usepackage{amsmath,amssymb,amsfonts}
\usepackage{algorithmic}
\usepackage{graphicx}
\usepackage{textcomp}
\usepackage{braket}
\usepackage{array}  % for better tables

\usepackage{cite}
\usepackage{hyperref}
\usepackage{xcolor}
\usepackage{stmaryrd}
\usepackage{quantikz}
\usepackage{booktabs}
\usepackage{float}
\usepackage{xcolor}
\ifx\noeditingmarks\undefined
    
\else
    
\fi

\ifx\noeditingmarks\undefined
    
\else
    
\fi

\newcommand{\para}[1]{\smallskip \noindent\textbf{#1}}

\newcommand{\cec}{\texttt{QRE-CEC}}

\newcommand{\none}{\texttt{QRE-None}}

\begin{document}

% \title{Comparing QEC-Enhanced 2G and 1G Quantum Repeaters: Performance Analysis Using SeQUeNCe}
\title{Realistic Simulation of Quantum Repeater with Encoding and Classical Error Correction}

\author{
  \IEEEauthorblockN{Sagar Patange$^*$, Caitao Zhan$^{\dagger}$, Bikun Li$^*$, Joaquin Chung$^{\dagger}$, Allen Zang$^*$, Liang Jiang$^*$, Rajkumar Kettimuthu$^{\dagger}$}
  \IEEEauthorblockA{$^{*}$The University of Chicago (USA), $^{\dagger}$Argonne National Laboratory (USA)}
  \IEEEauthorblockA{ sagarpatangev@gmail.com, czhan@anl.gov, bikunli@uchicago.edu, chungmiranda@anl.gov \\  yzang@uchicago.edu, liang.jiang@uchicago.edu, kettimuthu@anl.gov}
 
 \thanks{This material is based upon work supported by the U.S. Department of Energy, Office of Science, Advanced Scientific Computing Research (ASCR) program under contract number DE-AC02-06CH11357 as part of the InterQnet quantum networking project. }
 
 }

\maketitle

\begin{abstract}

Quantum repeaters are essential for scalable long-distance quantum networking. 
As quantum information processing moves toward fault-tolerant and error-corrected operations, it becomes increasingly important to study quantum repeaters that also move beyond raw physical entanglement and towards logical entanglement.
In this paper, we implement and simulate the quantum repeater with encoding and classical error correction (\cec) protocol in SeQUeNCe, a discrete-event simulator of quantum networks. 
The protocol distributes logical Bell pairs, performs encoded entanglement swapping, and uses classical error correction for the decoding of entanglement swapping measurement outcomes to determine Pauli-frame corrections.
For this study, we extend SeQUeNCe with a stabilizer-based backend, add support for CSS code-based encoded operations, and integrate gate, measurement, idle-decoherence, and state-initialization noise models. 
Our simulation results show that \cec\ suppresses all modeled errors to the second order. 
Also, \cec\ can distribute logical Bell pairs with 0.91 fidelity over a distance of 2000~km under the parameter regimes we study.
Beyond protocol-level performance evaluation, our implementation exposes practical simulator and control-plane challenges that are typically abstracted away in theoretical studies. 
% We release the implementation as open-source software.

\end{abstract}
\begin{IEEEkeywords}
Quantum Networks, Simulation, Logical Bell Pairs, Encoding, Error Correction, Stabilizer
\end{IEEEkeywords}

\section{Introduction}

Quantum networks~\cite{wehner2018quantum,rfc9340,interqnet-tqe-2026} extend quantum information processing beyond isolated devices, enabling long-distance quantum communication~\cite{duan-nature-2001,hu-teleportation-nrp23,quantum_repeaters-review2023}, distributed quantum computation~\cite{calaffi-dqc-cn24,dqc-nature25,switchqnet-isca25}, and networked quantum sensing~\cite{quantumsensing_2021,hillery-qsn-pra23,zhan-localization-qce23,zhan-qsn-tqc24,hu2025optimal,zang2026quantum}. 
These applications depend on distributing high-fidelity entanglement between distant nodes, which is limited by photon loss, imperfect operations, and finite memory lifetimes. 
Because direct optical-fiber transmission suffers exponential loss with distance, scalable quantum networks require quantum repeaters to divide long links into shorter segments and extend entanglement across them.

Quantum information processing is undergoing a paradigm shift~\cite{quantum_computing_cs_challenge_2026} from the Noisy Intermediate-Scale Quantum (NISQ) era towards early fault-tolerant and error corrected operation~\cite{google_qec-nature2023,acharya2025quantum,bluvstein2026fault}, where fragile physical qubits are increasingly replaced by encoded logical qubits~\cite{qec-guide-2019}. Early fault-tolerant quantum computation has been demonstrated in various qubit platforms, such as trapped-ion qubits~\cite{paetznick2024demonstration, reichardt2024tesseract}, superconducting qubits~\cite{acharya2025quantum, googlequantumai2025surface}, and neutral atoms~\cite{bluvstein2026fault}.
This raises a motivating question for quantum networks: should repeaters be capable of distributing logical Bell pairs protected by encoding and error correction rather than only physical ones? 
This question is especially important for long-distance entanglement distribution, where memory idling decoherence, latency, and imperfect operations compound across many links~\cite{oneway-prx-2020,zang2023entanglement,all_photonics-qce2024,hybrid-qcnc2024,mantri2025,hybrid-qce2025,repeater_graph_state-prl2025}.

% Quantum networks~\cite{wehner2018quantum} extend quantum information processing beyond isolated devices, enabling long-distance communication~\cite{quantum_repeaters-review2023}, distributed computation~\cite{dqc-nature25}, and networked sensing~\cite{zang2024quantum}. These applications depend on distributing high-fidelity entanglement between distant routers, which is limited by photon loss, imperfect operations, and finite quantum memory lifetimes. Because direct optical-fiber transmission suffers exponential photon loss with distance, and the quantum no-cloning theorem~\cite{wootters1982single} forbids amplifying quantum states as classical repeaters do with classical signals, scalable quantum networks rely on quantum repeaters, which break long links into shorter segments and extend entanglement across them.

% As quantum computing moves past the NISQ era toward early fault-tolerant operation~\cite{quantum_computing_cs_challenge_2026,google_qec-nature2023}, the same question arises for networks: should repeaters distribute encoded entanglement rather than just physical Bell pairs? The answer is especially relevant over long distances, where memory idling, control latency, and imperfect local operations compound errors across many links~\cite{hybrid-qcnc2024,mantri2025}.

Jiang et al.~\cite{jiangencoding-pra-2009} introduced a seminal quantum repeater architecture that combines CSS-code encoding with classical error correction to distribute encoded entanglement across long distances. We refer to this protocol as \textit{Quantum Repeater with Encoding and Classical Error Correction} (\cec) \cec. In \cec, neighboring stations generate logical Bell pairs, connect them through encoded entanglement swapping. 
Classical error correction is applied to the Bell state measurement outcomes to determine the Pauli-frame correction.
This architecture is particularly interesting because it represents a concrete shift from a physical-pair-based repeater toward a logical-pair-based repeater, and thus provides a useful model for studying how quantum networks may evolve toward more error-resilient operation.

Although \cec\ is well-defined theoretically, its practical behavior depends on system-level interactions that are difficult to evaluate analytically, including heralded entanglement generation, memory idling, encoded-state preparation, encoded swapping, classical error correction, and communication timing.
In this paper, we implement and simulate \cec\ in SeQUeNCe~\cite{sequence}, an open-source discrete-event simulator for quantum networks, to study how encoding and classical error correction affect long-distance entanglement generation under realistic timing and noise assumptions.
This work bridges the gap between theoretical proposals and realistic system-level implementation and evaluations.

\para{Contributions and Paper Organization}.
We define the network model and formulate the problem in \S\ref{sec:problem}. Our key contributions are:
\begin{enumerate}
\item We implement the \cec\ protocol proposed by Jiang et al.~\cite{jiangencoding-pra-2009} in the SeQUeNCe discrete-event quantum network simulator, adding the implementation parameters, system details, and simulator extensions needed for executable simulation (\S\ref{sec:protocol-design}, \S\ref{sec:implementation}). We release our implementation as open source\footnote{https://github.com/SagarPatange/Quantum-Repeater-Encoding}.

\item We identify hardware fidelity thresholds at which \cec\ achieves second-order error suppression across several network topologies, validating a key prediction of Jiang et al.~\cite{jiangencoding-pra-2009}.
Also, \cec\ can distribute logical Bell pairs with 0.91 fidelity over a distance of 2000~km under the parameter regimes we study (\S\ref{sec:results}).

\item Using the \cec\, we evaluate the ideal number of links needed to reach a target end-to-end logical fidelity at the desired distance under above error-threshold fidelity parameters (\S\ref{sec:results}).
\end{enumerate}
In \S\ref{sec:conclusion}, we conclude and discuss future work.

\section{Model, Problem and Related Work}
\label{sec:problem}

% 1) First define the network model
% 2) Formally present the problem statement 
% 3) Talk about the related work

% 1-2 Pages long

In this section, we define the quantum network model, the problem, and its benefits, and discuss related work.

\subsection{Network Model}

\begin{figure}[t]
    \centering
    \includegraphics[width=0.9\linewidth]{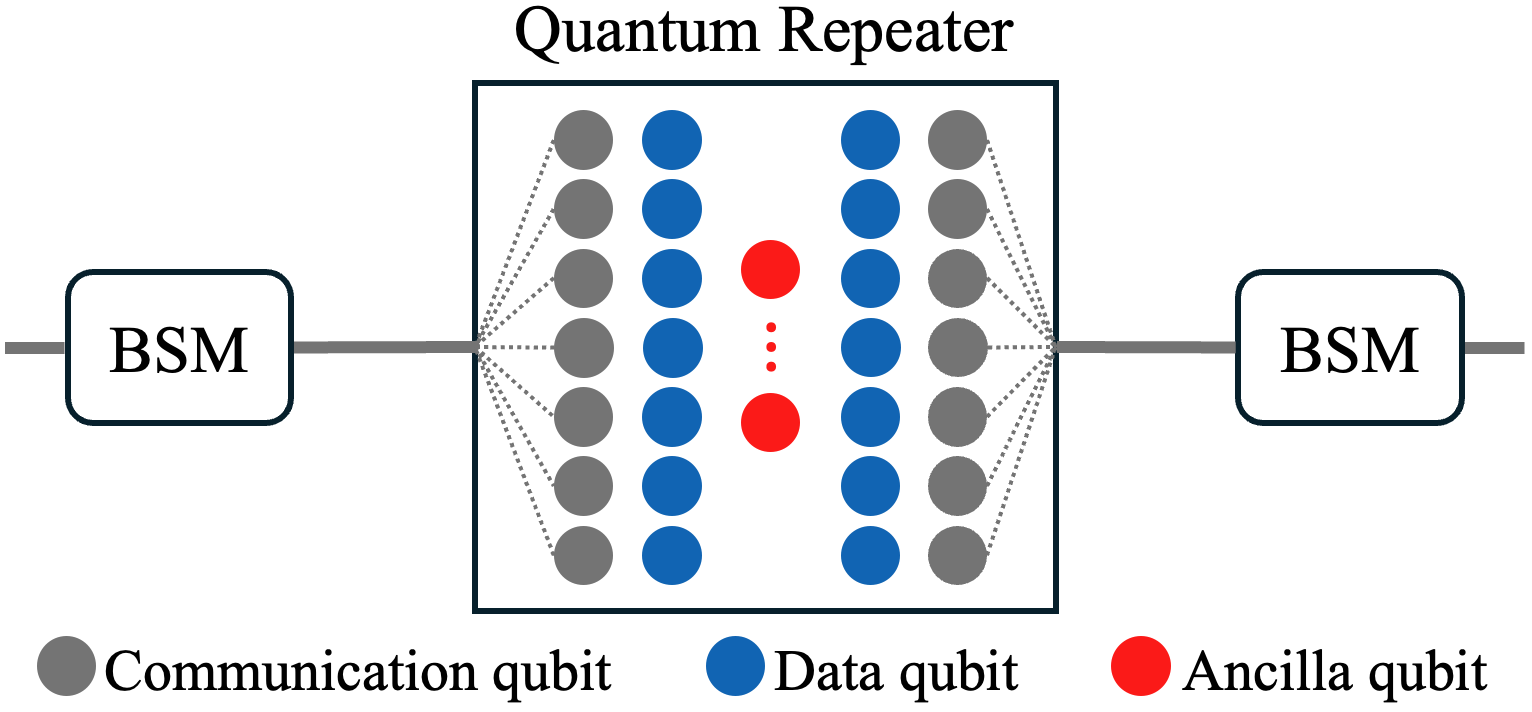}
    \vspace{-0.1in}
    \caption{A quantum repeater with encoding has three types of qubits: communication qubits (gray), data qubits (blue), and ancilla qubits (red).}
    \label{fig:repeater}
\end{figure}

We define a quantum network as a graph $G=(V, E)$, with $V=\{v_1, ..., v_N\}$ and $E=\{e(v_i, v_j) : v_i, v_j \in V\}$ denoting the set of nodes and links, respectively.
Pairs of nodes connected by a link are defined as neighbor nodes.
Each node is a quantum repeater that is composed of three types of qubits: communication qubits, data qubits, and ancilla qubits (see Fig.~\ref{fig:repeater}).

\begin{enumerate}
    \item The communication qubit (gray in Fig.~\ref{fig:repeater}) is used to interface with neighboring repeaters. It generates atom-photon entanglement and the photonic part is transmitted to a distant device for photonic Bell-state measurement (BSM). 
    
    \item The data qubit (blue in Fig.~\ref{fig:repeater}) is used to hold encoded states and logical Bell pairs.

    \item The ancilla qubits (red in Fig.~\ref{fig:repeater}) are helper qubits used during fault-tolerant encoded state preparation and syndrome extraction. While ancilla qubits assist with encoding operations, they are not part of the final encoded data block.
\end{enumerate}

The per-repeater qubit count is determined by the router's position in the topology and the protocol's chosen CSS $\llbracket n,k,d \rrbracket$ encoding type. $n$, $k$, $d$ represent the number of physical qubits required for encoding, the number of logical qubits encoded, and the code-distance, respectively. In our setup, we use the Steane $\llbracket 7,1,3 \rrbracket$ code~\cite{steane1996error}. Each Steane encoding block contains $n = 7$ data qubits, $n = 7$ communication qubits, and $1$ ancilla for our fault-tolerant state preparation scheme proposed by Goto~\cite{goto2016minimizing}. The number of encoding blocks a node contains is equal to the number of links connected to it. In a linear network topology, the edge nodes are responsible for 1 link and therefore contain 1 encoding block worth of qubits, while the middle nodes (quantum repeaters) require twice as many qubits since they are connected to 2 links, as illustrated in Figure~\ref{fig:repeater}.

Each link $e(u, v)$ in the quantum network is composed of two quantum channels, four classical channels, and a photonic BSM device in the middle.
The two quantum channels connect nodes $u$ and $v$ to the middle photonic BSM device, where photons emitted by the communication qubits are measured.
Two of the four classical channels connect the photonic BSM device to both $u$ and $v$ to send the measurement results,
while the other two classical channels connect $u$ to $v$ and vice versa to send messages related to protocol coordination.

 We assume the photons have three sources of loss (i) failure during the communication qubit emission, (ii) loss during the transmission in the quantum channel, and (iii) failure of the detector in the photonic BSM device. 
The qubits have five sources of noise (1) one-qubit gate error $p_{1g}$, (2) two-qubit gate error $p_{2g}$, (3) measurement error $p_m$, (4) idling decoherence error $p_{\mathrm{idle}}$, and (5) state initialization error $p_{\mathrm{init}}$. We assume the idling decoherence errors can be inserted as biased pauli channel noise, whos error coefficients $p_x^{\mathrm{idle}}$, $p_y^{\mathrm{idle}}$, and $p_z^{\mathrm{idle}}$ can be derived from $T_1$ and $T_2$ as shown in 
\S\ref{sec:implementation}.

\subsection{Problem}

This paper addresses the problem of realizing the \cec\ protocol~\cite{jiangencoding-pra-2009} in a discrete-event quantum network simulator and evaluating it under realistic network and hardware conditions. 
This problem is challenging because the protocol couples many effects that interact across both the quantum and classical planes. Heralded physical Bell-pair generation is probabilistic, so physical Bell pairs generated early will have to wait and incur idling decoherence. 
Encoded state preparation, teleported CNOT gates, encoded swapping, and classical error correction all introduce additional operations, synchronization points, and latency. As a result, end-to-end performance depends not only on gate and measurement fidelity, but also on memory coherence, control timing, and how the \emph{delays} and \emph{errors} accumulate throughout the protocol.
To support \cec, we extend SeQUeNCe~\cite{sequence} with a stabilizer-based simulation backend and error correction capabilities that it does not natively provide. These extensions let us simulate the full protocol stack, from heralded generation through Pauli-frame correction, under realistic timing and noise assumptions.

\subsection{Related Work}

Muralidharan et al.~\cite{muralidharan2016optimal} introduced the taxonomy of three generations of quantum repeaters. 
In that taxonomy, First-generation (1G) repeaters address photon loss through heralded entanglement generation and mitigate operational errors through heralded entanglement distillation. 
Second-generation (2G) repeaters replace heralded distillation with quantum error correction, eliminating the need for heralding in the correction of operational errors.
Third-generation (3G) repeaters go one step further by also replacing heralded entanglement generation with one-way transmission protected by quantum error correction, thereby addressing both loss and operational errors without the need for two-way classical signaling~\cite{inside-quantum-repeaters,ultrafast-prl-025}.
\cec\ is an early representative of 2G repeaters. It is characterized by using classical error correction rather than quantum error correction in the standard 2G repeaters.

% Muralidharan et al. compare all three generations across gate speed, fidelity, and coupling efficiency using a cost coefficient that captures the resources needed to achieve a target communication rate, identifying the parameter regimes where each generation is optimal.

% Several recent works extend the encoded repeater design space. Pathumsoot et al.~\cite{hybrid-qcnc2024} combine distillation with QEC in a hybrid strategy and show that distilled encoding outperforms either technique alone. 
% Mantri et al.~\cite{mantri2025} compare multiplexed two-way and one-way repeaters and find that two-way schemes remain competitive under updated memory availability assumptions. 
% Haldar et al.~\cite{hybrid-qce2025} study hybrid repeaters that mix different physical-memory platforms, analyzing encoded entanglement distribution with a three-qubit repetition code.
% Our work differs in focus. We focus on fully implementing the original \cec\ protocol in a discrete-event quantum network simulator to study the timing, noise, and control-plane interactions that arise in practice.
% Through realistic and extensive simulations, we display many insightful discoveries.

Several works have extended the repeater with encoding design space in different directions. 
Pathumsoot et al.~\cite{hybrid-qcnc2024} investigate hybrid error-management strategies that combine purification and quantum error correction, showing that purified encoding can improve end-to-end fidelity beyond what either technique achieves alone. 
Mantri et al.~\cite{mantri2025} compare multiplexed two-way and one-way repeaters under updated assumptions on memory availability and show that two-way schemes can remain competitive, or even superior, in relevant regimes.  
Haldar et al.~\cite{hybrid-qce2025} study hybrid repeaters that combine different physical-memory platforms and analyze encoded entanglement distribution using a three-qubit repetition code, emphasizing hardware heterogeneity and repeater placement.
In contrast, our work focuses on implementing the original \cec~\cite{jiangencoding-pra-2009} in a discrete-event simulator, with the goal of exposing the practical timing, noise, and control-plane issues that arise when implementing the protocol systematically.

\section{Protocol Design}
\label{sec:protocol-design}

\begin{figure}
    \centering
    \includegraphics[width=\linewidth]{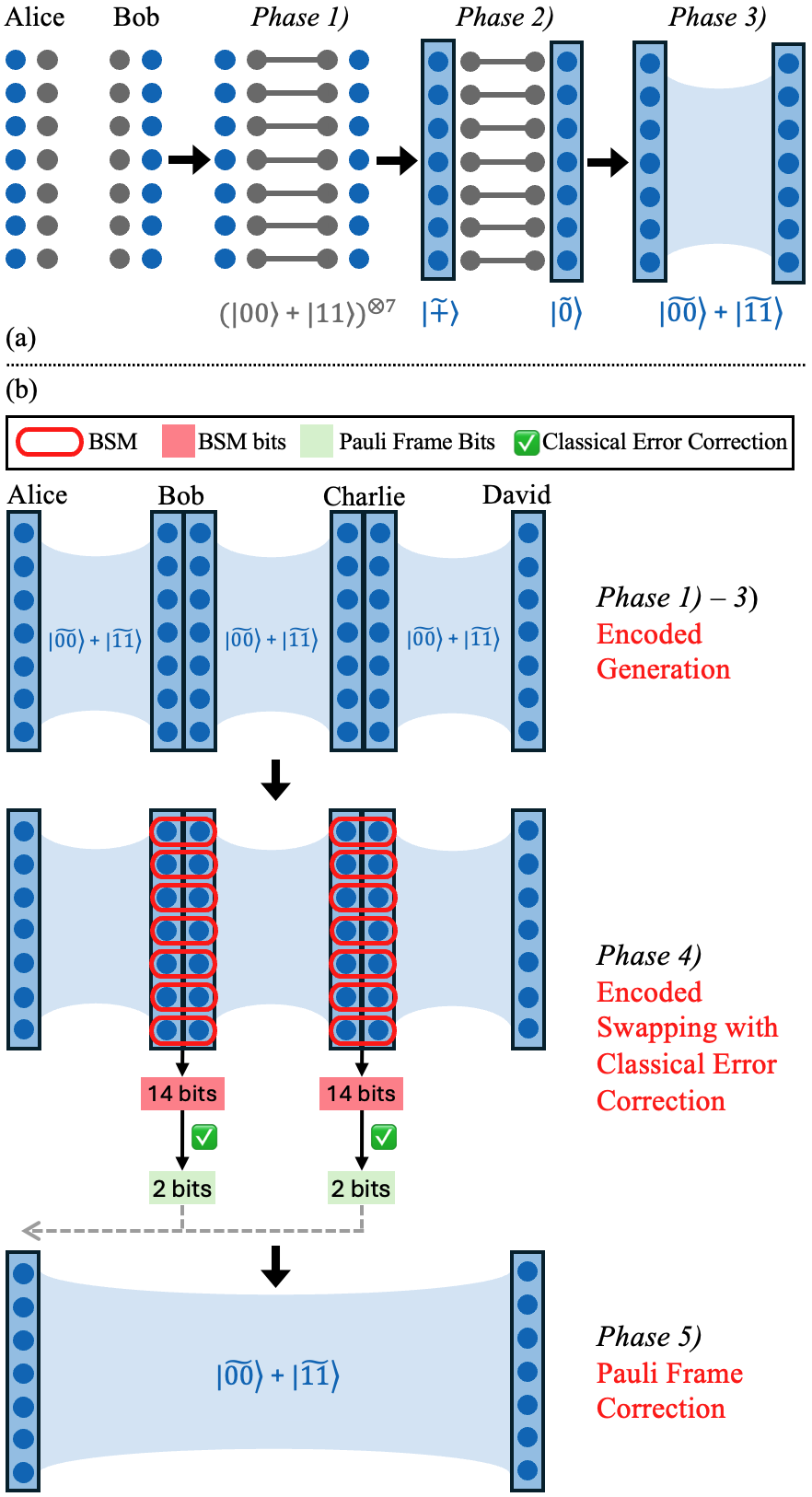}
    \vspace{-0.2in}
\caption{The \cec\ protocol. (a) Phases one through three: heralded entanglement generation, fault-tolerant logical state preparation, and logical teleported CNOT, which together produce encoded Bell pairs between neighboring nodes. (b) All five phases of \cec, adding encoded swapping with classical error correction and Pauli-frame correction to generate end-to-end logical Bell pairs.}
    \label{fig:protocol_overview}
\end{figure}

This section presents the design for our \textit{Quantum Repeater with Encoding and Classical Error Correction} (\cec) protocol.
The protocol has five phases (see Fig.~\ref{fig:protocol_overview}), and it can be generalized to any CSS code represented as $\llbracket n,k,d \rrbracket$. Specifically, Fig.~\ref{fig:protocol_overview} illustrates the Steane $\llbracket 7,1,3 \rrbracket$  code with $n = 7$.

% \begin{figure*}[t]
% \centering
% \includegraphics[width=\textwidth]{figures/protocol_overview.png}
% \caption{Overview of the repeater protocol with encoding for a four-repeater chain (Alice, Bob, Charlie, David). Step 1: encoded Bell pairs are generated between neighboring repeaters using physical Bell pairs and teleported CNOT. Step 2: middle repeaters (Bob, Charlie) perform encoded swapping via transversal CNOT and blockwise measurement, producing 14 raw bits each, which are classically decoded into 2 Pauli-frame bits. Step 3: the initiator (Alice) aggregates all frame bits and applies the final logical correction, yielding the end-to-end encoded Bell pair $|\widetilde{00}\rangle + |\widetilde{11}\rangle$ shared with David.}
% \label{fig:protocol_overview}
% \end{figure*}

\subsection{\cec's Five Phases}
Each logical-pair generation attempt runs inside a finite time window scheduled by the application layer. The protocol goes through five phases. A repeater only advances to the next phase once its barrier condition is met. These barrier conditions include readiness of other nodes, readiness of physical Bell pairs, and quantum memory state requirements to proceed to the next stage.

\subsubsection{Heralded Entanglement Generation (Barrett-Kok)}
Barrett-Kok is chosen for its ability to produce high-fidelity physical Bell pairs, as its two-round heralding scheme is robust to detector loss and spontaneous emission~\cite{barrett2005efficient}. 
At the midpoint BSM station, photons emitted by neighboring repeaters interfere and are measured by single-photon detectors. Successful heralding across both rounds confirms an entangled state ($|\Phi^+\rangle$). 
Memory efficiency, fiber attenuation, and detector efficiency contribute to link-level physical Bell pair generation latency.

\subsubsection{Fault-Tolerant Logical State Preparation}
\begin{figure}[ht]
\centering
\begin{quantikz}[row sep=0.1cm, column sep=0.1cm]
\lstick{$|0\rangle$}  & \qw       & \targ{}   & \qw       & \qw       & \targ{}   & \qw       & \qw       & \ctrl{7} & \qw      & \rstick[wires=7]{$|\tilde{0}\rangle$} \qw \\
\lstick{$|0\rangle$}  & \gate{H}  & \ctrl{-1} & \qw       & \ctrl{3}  & \qw       & \ctrl{4}  & \qw       & \qw      & \qw      & \qw      \\
\lstick{$|0\rangle$}  & \gate{H}  & \qw       & \ctrl{4}  & \qw       & \ctrl{-2} & \qw       & \qw       & \qw      & \qw      & \qw      \\
\lstick{$|0\rangle$}  & \gate{H}  & \ctrl{2}  & \qw       & \qw       & \ctrl{3}  & \qw       & \qw       & \qw      & \qw      & \qw      \\
\lstick{$|0\rangle$}  & \qw       & \qw       & \qw       & \targ{}   & \qw       & \qw       & \targ{}   & \qw      & \qw      & \qw      \\
\lstick{$|0\rangle$}  & \qw       & \targ{}   & \qw       & \qw       & \qw       & \targ{}   & \qw       & \qw      & \ctrl{2} & \qw      \\
\lstick{$|0\rangle$}  & \qw       & \qw       & \targ{}   & \qw       & \targ{}   & \qw       & \ctrl{-2} & \qw      & \qw      & \ctrl{1} \\
\lstick{$|0\rangle$}  & \qw       & \qw       & \qw       & \qw       & \qw       & \qw       & \qw       & \targ{}  & \targ{}  & \targ{}  & \qw    & \meter{M_{0/1}}
\end{quantikz}
\caption{Paetznick--Reichardt 8-CNOT encoder with minimal verification~\cite{goto2016minimizing}. 
Qubits 1--7 are the data block encoding $|\tilde{0}\rangle$. The ancilla (bottom) 
measures $ZIIIIZZ$; a non-trivial syndrome triggers block rejection and re-preparation.}
\label{fig:ft_encoding}
\end{figure}
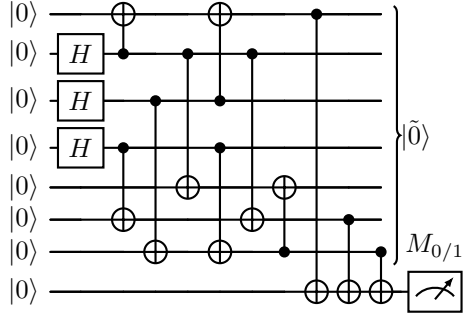

Once a link (a pair of neighboring nodes) holds $n$ physical Bell pairs, data qubits on the control block are encoded into the logical $|\tilde{+}\rangle$ and the target block encoded into the logical $|\tilde{0}\rangle$ state. For this logical encoding we use the Steane 
$\llbracket 7,1,3 \rrbracket$ code via the Paetznick--Reichardt 8-CNOT 
encoder~\cite{paetznick2013fault}, which reduces the CNOT count 
relative to conventional 9- or 12-CNOT encoders and lowers circuit noise. 
Goto's scheme~\cite{goto2016minimizing} further improves fault tolerance by introducing a flag qubit that measures the weight-3 logical-$\bar{Z}$ verification operator $ZIIIIZZ$. A non-trivial measurement outcome triggers block rejection and re-preparation, reducing the probability of accepting a corrupted encoded block. The encoding circuit with verification is shown in Fig.~\ref{fig:ft_encoding}.

\subsubsection{Transversal Teleported CNOT}
\begin{figure}[ht]
\centering
\begin{quantikz}[row sep=0.15cm, column sep=0.4cm]
\lstick{$d_A: | \alpha \rangle$}      & \qw & \ctrl{1}  & \qw              & \gate{Z}           & \rstick{$| \alpha \rangle$} \qw \\
\lstick{$c_A\vphantom{|}$} & \lstick[wires=2]{\scriptsize $|\Phi^+\rangle$}\qw & \targ{}  & \meterD{Z}\vcw{2} \\
\lstick{$c_B\vphantom{|}$} &                                    \qw & \ctrl{1} & \qw & \meterD{X}\vcw{-2} \\
\lstick{$d_B: |\beta\rangle$}  & \qw & \targ{}   & \gate{X}         & \qw                & \rstick{$|{\beta} \oplus \alpha \rangle$} \qw
\end{quantikz}
\caption{Gate teleportation circuit for the nonlocal CNOT~\cite{jiang2007distributed}. A
Bell pair $|\Phi^+\rangle$ is shared between communication
qubits $c_A$ and $c_B$. Alice applies a local CNOT from
her data qubit $d_A$ to $c_A$ and measures $c_A$ in the
$Z$ basis. Bob applies a local CNOT from $c_B$ to his data
qubit $d_B$ and measures $c_B$ in the $X$ basis. The
measurement outcomes determine conditional Pauli corrections:
$Z$ on the control ($d_A$) from Bob's result and $X$ on the target ($d_B$) from
Alice's result.}
\label{fig:teleported_cnot}
\end{figure}
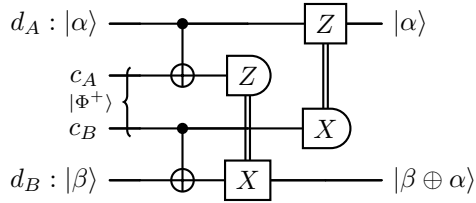

Once encoding is completed on both sides of a link, the two neighboring repeaters run $n = 7$ transversal teleported CNOTs following the gate teleportation circuit~\cite{jiang2007distributed}, shown in Fig.~\ref{fig:teleported_cnot}. Each repeater applies a local CNOT between its data and communication qubits, then measures the communication qubit in the appropriate basis depending on its role (Alice or Bob). The two repeaters exchange measurement outcomes and each applies a conditional Pauli correction to its local encoded block based on the other's result. All qubits idle and decohere while waiting for the classical messages. 
After corrections, the two encoded blocks share logical entanglement, and the $n$ shared Bell pairs are consumed. Because the CNOT is applied transversally, a single-qubit error on one block cannot spread to more than one qubit on the other, so the step is fault-tolerant by construction. Single-qubit errors are therefore correctable by the Steane code, and the logical error rate is suppressed to the $2^{nd}$ order. A middle repeater does not advance to encoded swapping until this step has finished on both its adjacent links.

\subsubsection{Encoded Swapping with Classical Error Correction}
The encoded swapping stage only starts when the transversal teleported CNOT has been successful on both sides of the quantum repeater, i.e., after the quantum repeater carries two logical Bell pairs. At each intermediate repeater, encoded entanglement swapping extends the logical 
entanglement across the chain. The repeater applies a transversal CNOT between 
its left and right encoded blocks, then measures the left block in the $X$ basis 
and the right block in the $Z$ basis, producing two raw bit strings 
$\mathbf{m}_x, \mathbf{m}_z \in \{0,1\}^7$ (treated as column vectors). For each string, the classical
decoder computes the 3-bit syndrome
\begin{equation}
  \mathbf{s} = H \mathbf{m} \bmod 2
  \label{eq:syndrome}
\end{equation}
where $H$ is the parity-check matrix of the classical $[7,4,3]$ Hamming code~\cite{steane1996error} underlying the Steane $\llbracket 7,1,3 \rrbracket$ code,
\begin{equation}
  H = \begin{pmatrix}
    0 & 0 & 0 & 1 & 1 & 1 & 1 \\
    0 & 1 & 1 & 0 & 0 & 1 & 1 \\
    1 & 0 & 1 & 0 & 1 & 0 & 1
  \end{pmatrix}
  \label{eq:hamming_parity}
\end{equation}
Because the Steane code is a CSS code, $X$ and $Z$ errors produce independent 
syndromes and are decoded separately using the same Hamming 
decoder~\cite{jiangencoding-pra-2009, qcqi-book}. The columns of $H$ are the 
binary representations of 1 through 7, so a nonzero syndrome directly identifies 
the error location (see Table~\ref{tab:syndrome_decode}). The decoder flips the indicated bit of the raw outcome string $\mathbf{m}$ to obtain the corrected string $\mathbf{m}'$. A zero syndrome leaves $\mathbf{m}' = \mathbf{m}$. The logical frame bits are then extracted from $\mathbf{m}'$ by computing
the parity across all seven positions:
\begin{equation}
  b_x = \bigoplus_{i=1}^{7} (\mathbf{m}'_x)_i \qquad \ \
  b_z = \bigoplus_{i=1}^{7} (\mathbf{m}'_z)_i
  \label{eq:frame_extract}
\end{equation}
These constitute the Pauli-frame contribution from that repeater. The frame bits 
$b_x$ and $b_z$ from all intermediate repeaters are forwarded classically to the 
end nodes, where they are combined to determine the total Pauli frame correction 
required to recover the logical Bell pair.
\begin{table}[t]
\renewcommand{\arraystretch}{1.3}
\centering
\caption{Steane syndrome decoding table.}
\label{tab:syndrome_decode}
\footnotesize
\begin{tabular}{|c|c|}
\hline
\textbf{Syndrome $\mathbf{s}$} & \textbf{Error location} \\
\hline
000 & None \\
001 & Qubit 1 \\
010 & Qubit 2 \\
011 & Qubit 3 \\
100 & Qubit 4 \\
101 & Qubit 5 \\
110 & Qubit 6 \\
111 & Qubit 7 \\
\hline
\end{tabular}
\end{table}

\subsubsection{Pauli-Frame Correction}
The two decoded bits from each middle repeater form a Pauli-frame contribution $(b_x, b_z)$, which tells the initiator how the current end-to-end state differs from the target Bell pair. Each middle node sends its two bits to the initiator over a classical channel. No quantum information is transmitted at this stage.
The initiator waits for all $N-2$ middle repeaters to report in, where $N$ is the length of the repeater chain, then aggregates the frame bits by XOR:

\begin{equation}
  b_x^{\mathrm{final}} = \bigoplus_{k=1}^{N-2} b_x^{(k)}
  \qquad \ \ 
  b_z^{\mathrm{final}} = \bigoplus_{k=1}^{N-2} b_z^{(k)}
  \label{eq:frame_agg}
\end{equation}
The resulting correction is applied as a logical $\tilde{X}$ and $\tilde{Z}$ on the initiator's local encoded block. The entire frame is deferred to any endpoint repeater, which in our case was the initiator. This one-way classical communication pattern is the key difference between second-generation repeater architectures~\cite{jiangencoding-pra-2009} from first-generation protocols, which require multiple rounds of two-way signaling for entanglement purification~\cite{bbpssw}.

Once the frame correction is applied, the initiator evaluates the end-to-end
logical fidelity after ideal recovery of the two endpoint encoded blocks, as
described in Section~\ref{sec:fidelity}. This recovery step removes correctable
physical errors before the final projection onto the target logical Bell pair,
so the reported metric captures logical rather than physical fidelity. The
fidelity is then computed from the logical correlators
$\langle \tilde{X}\tilde{X} \rangle$,
$\langle \tilde{Y}\tilde{Y} \rangle$, and
$\langle \tilde{Z}\tilde{Z} \rangle$. 
% This fidelity and the total protocol latency are the two primary metrics recorded for each run.

% Is how we implement the Tableau - anything relating to SeQUeNCe. The extension 

\section{Implementation in SeQUeNCe}
\label{sec:implementation}

\begin{figure}
    \centering
    \includegraphics[width=\linewidth]{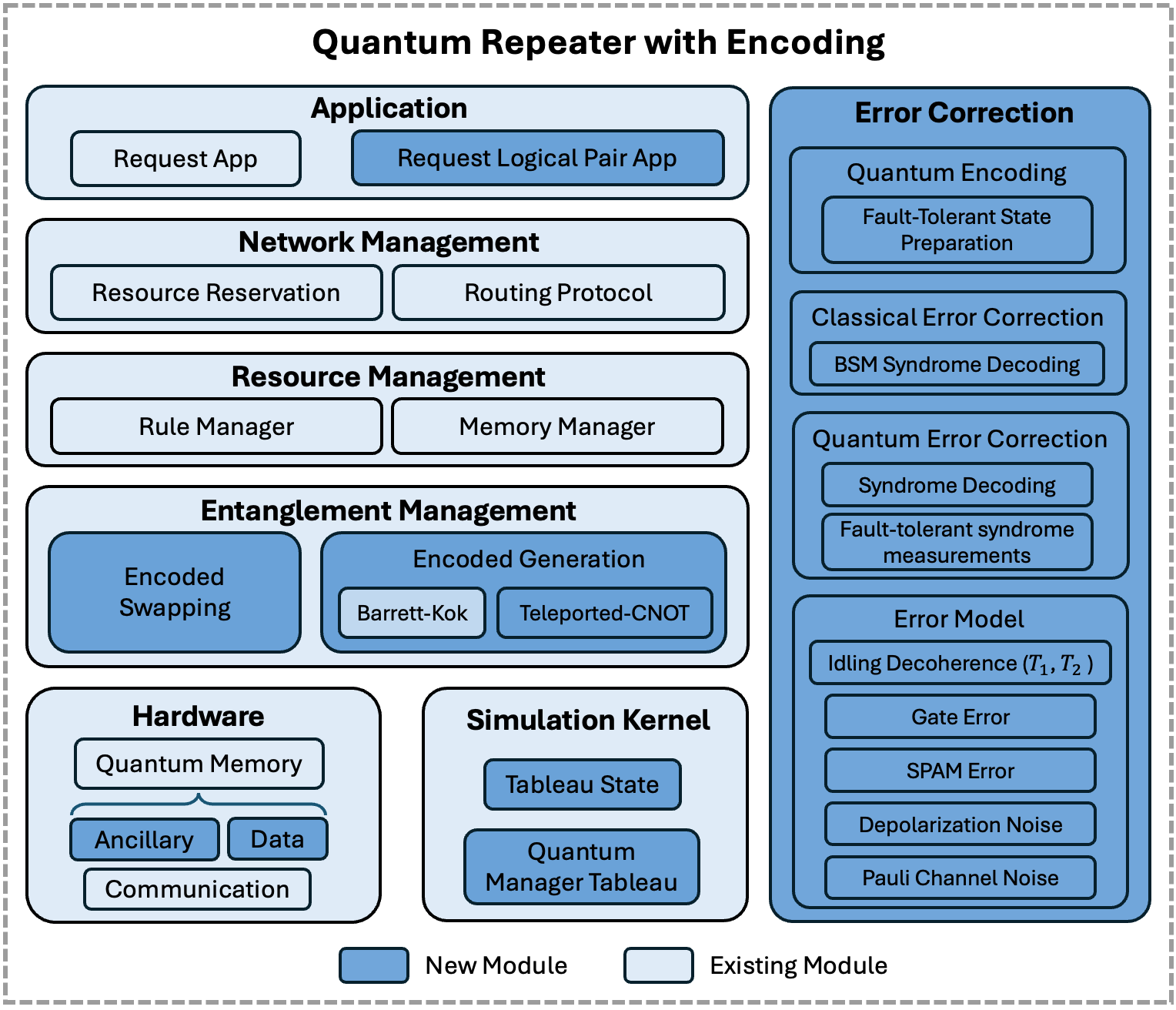}
    \vspace{-0.2in}
    \caption{Illustration of new modules introduced to SeQUeNCe to fully support the implementation of a quantum repeater with encoding and error correction.}
    \label{fig:sequence}
\end{figure}

In recent years, several quantum network simulators have been developed, such as SeQUeNCe~\cite{sequence}, QuISP~\cite{quisp}, NetSquid~\cite{netsquid}, and QuantumSavory~\cite{quantumsavory}.
We chose SeQUeNCe for our simulation because it is open-source, well-documented, and easy to extend~\cite{zang2022simulation,transducer-qce-2025,qlan-icc-2025,hetero-qcnc-2026}. 
SeQUeNCe is organized into six modules: 1) Application, 2) Network Management, 3) Resource Management, 4) Entanglement Management, 5) Hardware, and 6) Simulation Kernel. 
See Fig.~\ref{fig:sequence}. 
This section describes the extensions to SeQUeNCe we make to support the \cec\  protocol. 

% The logical operations in our protocol (encoding, transversal
% CNOT, stabilizer measurement, Pauli correction) are all
% Clifford operations. We use the tableau formalism to avoid
% the exponential cost of state-vector simulation, with Stim's
% \texttt{TableauSimulator}~\cite{gidney2021stim} as the
% quantum state backend for all QEC experiments.

% \subsection{SeQUeNCe Overview}

% Requests originate in the Application module and are passed to the Resource Reservation Protocol, which installs rules in the Resource Manager. Once a rule's condition is met, it spawns an entanglement protocol tied to a specific set of quantum memories on the node.

% \subsection{Code Extensions}

\paragraph*{1) Quantum Manager with Tableau Backend}
SeQUeNCe's existing backend uses ket-vectors and density matrices, whose memory scales with $O(2^n)$ and $O(2^{2n})$ respectively. 
This makes them computationally intractable beyond the representation of 10-20 qubits, which is far below the resource requirements of long-distance encoded quantum repeater chains that require hundreds of qubits. Memory in the tableau formalism scales as $O(n^2)$, enabling the simulation of thousands of qubits, with the caveat that only Clifford operations can be simulated efficiently. This limitation is not an issue, however, since the majority of quantum network protocols only contain Clifford operations. To efficiently represent the Clifford operations we add \texttt{QuantumManagerTableau}, a new backend built on Stim~\cite{gidney2021stim, aaronson2004improved}. It utilizes a Tableau to represent qubits as stabilizer states with phases as shown in Table~\ref{tab:psi_plus_tableau}.

\begin{table}[h]
\centering
\vspace{-0.1in}
\caption{Stabilizer tableau for $|\Phi^+\rangle$}
\label{tab:psi_plus_tableau}
\begin{tabular}{c|cc|cc|c}
\hline
 Generator & $x_1$ & $x_2$ & $z_1$ & $z_2$ & $r$ \\
\hline
$XX$ & 1 & 1 & 0 & 0 & + \\
$ZZ$ & 0 & 0 & 1 & 1 & + \\
\hline
\end{tabular}
\end{table}

Each entangled group of qubits is tracked in a \texttt{TableauState} object pairing a \texttt{TableauSimulator} object from stim with the SeQUeNCe memory keys sharing that state; unentangled qubits occupy separate objects. When an operation requires joint access across groups, such as a transversal CNOT between encoded blocks at neighboring routers, the relevant \texttt{TableauState} objects are merged, and then gates are applied.
Clifford gates are executed directly in Stim, with gate noise injected immediately afterward. Upon measurement, measured qubits are decoupled from the active stabilizer tableau to keep individual tableaus small and computationally efficient.

\paragraph*{2) Quantum Memory}
The \texttt{Memory} class is extended to distinguish communication, data, and ancilla roles. Communication memories handle heralded Bell-pair generation, data memories store the encoded block, and ancilla memories are used during fault-tolerant state preparation. The manager tracks \texttt{last\_idle\_time} per qubit so idling decoherence can be applied before each major phase.

\paragraph*{3) Fault-Tolerant State Preparation}
Fault-tolerant state preparation in our implementation is exposed through three code-level modes: \texttt{none}, \texttt{minimal}, and \texttt{standard}. All three use the same Paetznick--Reichardt-style 8-CNOT Steane encoder, implemented in the codebase as part of the Steane \texttt{CSSCode} logic. In \texttt{none}, the block is accepted immediately after encoding. In \texttt{minimal}, the implementation performs one Steane verification check using a single ancilla. In \texttt{standard}, it performs four verification checks. Any nonzero verification outcome causes the encoded block to be reset and re-prepared.

\paragraph*{4) Teleported Logical CNOT}
The \texttt{TeleportedCNOT}-\texttt{Protocol} class, along with \texttt{TeleportedCNOTMessage} and \texttt{TeleportedCNOTMsgType}, implements the teleported logical CNOT. For a given neighboring link, the protocol runs the two-party control flow over local data and communication qubits, consuming $n = 7$ physical Bell pairs in parallel. It is organized into Alice/Bob phases, with explicit classical messages carrying each side's measurement results and a final completion checkpoint. Classical communication delays are modeled, and routers buffer incoming messages if they are not yet ready to apply corrections or act on the message instructions. Since the teleported CNOT is applied transversally between two encoded blocks, it is fault tolerant by construction because a single-qubit error on one side cannot spread to more than one qubit on the other.

\paragraph*{5) Encoded Swapping}
The \texttt{QREProtocol} class handles encoded swapping at middle routers. It only runs once both adjacent teleported-CNOT links have completed. The protocol collects the router's left and right encoded blocks, applies transversal CNOTs between them, then measures the left block in the $X$ basis and the right block in the $Z$ basis. The resulting raw Steane-block outcomes are decoded into Pauli-frame contributions for the end-to-end logical Bell pair.

\paragraph*{6) Classical Error Correction}
Classical error correction uses the \texttt{QREMessage} and \texttt{QREMsgType} messaging layer together with the Steane decoding routines in \texttt{CSSCode} and \texttt{Steane713}. The decoder computes separate 3-bit syndromes from the left-block $X$-measurement bits and the right-block $Z$-measurement bits, flips at most one bit per side in \texttt{cec} mode, and extracts the corrected logical parity. Each middle router packs its frame contributions into a \texttt{QREMessage} and sends it to the initiator, where \texttt{QREProtocol} XORs them together to get the final logical $\tilde{X}$ and $\tilde{Z}$ frame corrections.

\paragraph*{7) Noise Model}
% The simulator has three noise mechanisms. 
Gate noise is modeled as a depolarizing channel applied after each ideal Clifford gate. 
% Depolarizing noise is the worst case because it weights all Pauli errors equally, so the fidelities we report are a lower bound. 
The simulator also supports a general biased Pauli channel for gate noise (\texttt{PAULI\_CHANNEL\_1} for one-qubit gates and \texttt{PAULI\_CHANNEL\_2} for two-qubit gates), where the total error rate is distributed across the $\{X, Y, Z\}$ axes according to a bias vector $(w_x, w_y, w_z)$, and correlated two-qubit errors such as $XX$ or $YX$ are weighted to occur less frequently than single-qubit errors like $IX$ or $YI$, which is a more realistic noise model. We do not use the biased channel in the results presented here. 
One-qubit gates use $p = \tfrac{3}{2}(1 - F_{1q})$ applied as \texttt{DEPOLARIZE\_1}, and two-qubit gates use $p = \tfrac{5}{4}(1 - F_{2q})$ applied as \texttt{DEPOLARIZE\_2}. Measurement noise flips the reported outcome with probability $1 - F_{\mathrm{meas}}$. 
For idling decoherence, a depolarizing channel does not work because it cannot capture the asymmetry between $T_1$ and $T_2$ processes. We use a biased Pauli channel (\texttt{PAULI\_CHANNEL\_1}) on each idle qubit instead, with error probabilities $p_x^{\mathrm{idle}}$, $p_y^{\mathrm{idle}}$, and $p_z^{\mathrm{idle}}$ derived from $T_1$ and $T_2$:
\begin{equation}
\label{eq:idle}
\begin{aligned}
p_x^{\mathrm{idle}} &= p_y^{\mathrm{idle}} = \frac{1 - e^{-t/T_1}}{4} \\
p_z^{\mathrm{idle}} &= \frac{1 - e^{-t/T_2}}{2} - \frac{1 - e^{-t/T_1}}{4}
\end{aligned}
\end{equation}
where $t$ is the idle time of the qubit. The constraint $2T_1 \geq T_2$ ensures $p_z^{\mathrm{idle}} \geq 0$: pure dephasing contributes only to $T_2$, while $T_1$ processes contribute to both, so $T_2$ cannot exceed $2T_1$. The Pauli channel model is valid assuming the channels have undergone Pauli twirling~\cite{ghosh2012surface}. Bell-pair infidelity is injected at the physical layer during heralded generation as it flips the final Bell state into one of the 4 Bell state. State initialization noise represents errors when preparing the $|0\rangle$ state.

\paragraph*{8) Fidelity Extraction}
\label{sec:fidelity}
End-to-end logical fidelity is computed at the initiator after the final Pauli-frame update. Because the encoded protocol can correct small physical errors, we apply an ideal recovery operation $\mathcal{R}$ to the stabilizer tableau of each endpoint block before evaluating logical fidelity. $\mathcal{R}$ is the noiseless Steane decoder: syndrome extraction followed by the inferred single-qubit Pauli correction, applied independently to each endpoint block. 
Without this step, even a single weight-one physical error would move the state outside the code space and show up as infidelity, even though the Steane code can correct it.
Given the density matrix of the final state $\mathcal{R}(\rho_{\mathrm{final}})$, the end-to-end logical fidelity is computed by $F = \langle \tilde{\Phi}^+ | \mathcal{R}(\rho_{\mathrm{final}}) | \tilde{\Phi}^+ \rangle$.
In the stabilizer formalism, however, this is equivalently computed by measuring the expectation values of the logical Bell-state stabilizers directly on the 
\texttt{TableauSimulator} state using 
\texttt{peek\_observable\_expectation}, giving
\begin{equation}
\label{eq:bell_correlator_fidelity}
F =
\frac{
1
+ \langle \tilde{X}\tilde{X} \rangle
- \langle \tilde{Y}\tilde{Y} \rangle
+ \langle \tilde{Z}\tilde{Z} \rangle
}{4}
\end{equation}
Because \texttt{stim} evolves one sampled stabilizer trajectory at a 
time, each run gives an exact trajectory-level fidelity. The results in
Section~\ref{sec:results} report the average over tens of thousands of independent runs for reasonable precision.

\paragraph*{9) Router and Topology}
The implementation uses a \texttt{QuantumRouter2ndGeneration} node type, which extends SeQUeNCe's base \texttt{RuantumRouter} class with separate communication, data, and ancilla memory arrays. 
A corresponding \texttt{RouterNetTopo2G} topology class instantiates these routers from JSON configuration files, expands shared memory templates onto the data and ancilla arrays, and propagates node-level hardware and protocol settings such as gate fidelity, measurement fidelity, two-qubit gate fidelity, characteristic $T_1$ and $T_2$ times, state-initialization fidelity, and fault-tolerant preparation mode into each quantum router.

\section{Simulation Results}
\label{sec:results}

In this section we present simulation results demonstrating the superior end-to-end fidelity of \cec\ over bare \none\ (no classical error correction) for encoded logical Bell pair generation.

\subsection{Simulation Settings}
In our simulation, we discuss four categories of parameters: network-scale parameters (number of repeaters, inter-repeater distance), hardware-noise parameters (two-qubit gate fidelity, one-qubit gate fidelity, measurement fidelity, state initialization fidelity, dephasing $T_2$ time, and energy relaxation $T_1$ time), protocol-level parameters (fault-tolerant preparation mode, classical correction modes), and code-level parameters (CSS codes). The primary metrics measured through selective parameter sweeps are end-to-end logical Bell-pair fidelity and latency. Table~\ref{tab:simulation_parameters} summarizes the key parameters for our simulation.

\begin{table}[t]
\caption{Simulation parameters and notations}
\label{tab:simulation_parameters}
\centering
\fontsize{8}{7}\selectfont
\renewcommand{\arraystretch}{1.8}
\setlength{\tabcolsep}{4pt}
\begin{tabular}{|p{0.1\columnwidth}|p{0.28\columnwidth}||p{0.1\columnwidth}|p{0.35\columnwidth}|}
\hline
\textbf{Symbol} & \textbf{Meaning} & \textbf{Symbol} & \textbf{Meaning} \\
\hline\hline
$F_{\mathrm{init}}$ & Initialization fidelity & $F_m$ & Measurement fidelity \\
\hline
$F_{1q}$ & 1-Qubit gate fidelity & $T_1$ & Amplitude-damping time \\
\hline
$F_{2q}$ & 2-Qubit gate fidelity & $T_2$ & Dephasing time \\
\hline
$\alpha$ & Fiber attenuation   & $F_{\mathrm{phys}}$ & Physical Bell-pair fidelity  \\
\hline
$c^*$ & Speed of light in fiber & $\eta_d$ & Photon detector efficiency \\
\hline
$\eta_m$ & Memory efficiency & $z$ & Hardware sweep parameter \\
\hline
$D_{\mathrm{end}}$ & End-node local delay & $D_{\mathrm{fwd}}$ &  Forwarding delay \\
\hline
\end{tabular}
\end{table}

\subsubsection{Network Topology}
We consider linear network topologies with a BSM node at the midpoint of each elementary link. The number of nodes and their spacing pull in opposite directions. Adding more intermediate repeaters within a fixed total distance cuts per-link latency because photon loss drops exponentially with shorter links, but it also deepens the circuit and accumulates more gate and measurement errors. Spacing repeaters further apart keeps the circuit shallow but increases idle time on all qubits while waiting for photons, requiring higher $T_1$ and $T_2$ to compensate. We sweep both the number of links and the inter-node distance in our results to explore this.

\subsubsection{Classical Communication Latency} Classical communication carries 
measurement outcomes and coordination messages.
The latency~\cite{acp-qcnc-2025} between 
nodes $u$ and $v$ is:
\begin{equation}
    l_{(u,v)} = \frac{d_{(u,v)}}{c^*} + \text{hop}_{(u,v)} \times D_{\mathrm{fwd}} + D_{\mathrm{end}}
\end{equation}
where $d_{(u,v)}$ is the path length, $c^* = 2\times10^{8}~\mathrm{m/s}$,
$D_{\mathrm{fwd}} = 20~\mu\mathrm{s}$ is the per-hop forwarding delay, and
$D_{\mathrm{end}} = 50~\mu\mathrm{s}$ is the end-node processing delay.
Transmission and queueing delays are neglected.

\subsubsection{Hardware Settings}
The key latency-driving parameters are quantum memory efficiency $\eta_m = 0.90$, 
photon detector efficiency $\eta_d = 0.95$, and fiber attenuation 
$\alpha = 0.2~\mathrm{dB/km}$. 
% The network has $1$, $2$, or $5$ links at $20~\mathrm{km}$ each.
These values are somewhat optimistic relative 
to typical experimental demonstrations today, representing near-term targets: 
SNSPD efficiencies above $93\%$ at telecom wavelengths have been 
reported~\cite{marsili2013detecting, zadeh2021snspd}, and memory efficiencies 
approaching $70$--$80\%$ have been demonstrated in telecom-compatible rare-earth 
and solid-state platforms~\cite{knaut2024entanglement}, though $90\%$ remains a 
near-term target. Fiber parameters follow standard telecom specifications~\cite{sequence}.

\subsubsection{Noise Settings}
\begin{figure*}[h]
    \includegraphics[width=\linewidth]{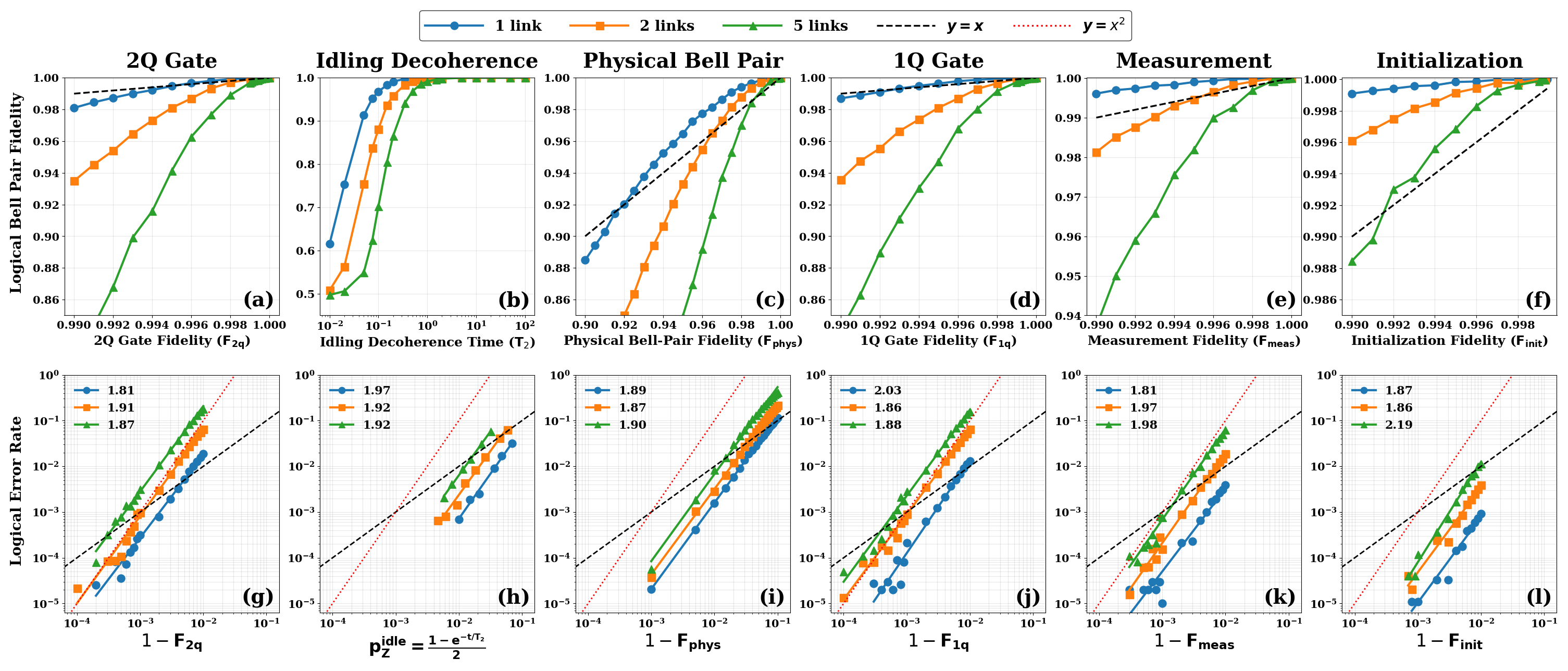}
    \vspace{-0.3in}
\caption{
    Logical Bell-pair fidelity under sweeps of six physical error parameters for linear topologies with $1$, $2$, and $5$ links with 20~km link distance.
    In each sweep, all other noise sources are set to zero.
    Subfigures (a)--(f) show the logical Bell-pair fidelity as a function of two-qubit gate fidelity, idling decoherence time $T_2$, physical Bell-pair fidelity, one-qubit gate fidelity, measurement fidelity, and initialization fidelity, respectively.
    In (a)--(f), the $y = x$ line marks the threshold: points above it mean the end-to-end logical fidelity exceeds the physical fidelity parameter being swept.
    Subfigures (g)--(l) show the corresponding logical error rate versus the associated physical error parameter on log--log axes, with fitted scaling exponents indicated by the solid lines.
    The dashed black line denotes $y = x$, while the dotted red line provides a slope-$2$ scaling reference corresponding to shifted $y \propto x^2$ behavior.
    The slopes in legends (g)--(l) are close to $2$, confirming second-order error suppression.}
    \vspace{-0.2in}
    \label{fig:thresholds}
\end{figure*}

Gate noise is modeled as a depolarizing channel as described in the noise model. Gate fidelities and coherence times used in the sweeps are based on recent neutral-atom experiments~\cite{evered2025benchmarking, oxford2025singlequbit, bluvstein2026fault, manetsch2025tweezer}. We use the Steane $\llbracket 7,1,3 \rrbracket$ code~\cite{steane1996error} with fault-tolerant state preparation in minimal mode~\cite{goto2016minimizing}.
In~\cite{jiangencoding-pra-2009}, the effective error probability per physical qubit is $q = 4\beta + 2\mu + \delta$, where $\beta$ is the gate error probability, $\mu$ is the memory error probability, and $\delta$ is the measurement error probability. 
The logical error probability scales as $Q \sim q^{t+1}$ for an $\llbracket n, k, 2t+1 \rrbracket$ CSS code. 
Here $t$ is the maximum number of arbitrary physical-qubit errors that a code can correct, and the code distance $d=2t+1$.
Defining a single $q$ in our case is hard because $T_1$ and $T_2$ decoherence is time-dependent and accumulates differently depending on qubit idle times, gates and measurements happen at variable points during the teleported CNOT, and initialization infidelity enters separately. Instead of collapsing all of this into one number, we use a parameter $z \in [0,1]$ that interpolates each noise parameter between a baseline value ($z=0$) and the ideal limit ($z=1$). For fidelity parameters the interpolation is linear,
\begin{equation}
    F(z) = F_0 + (1 - F_0)\,z
    \label{eq:z_fidelity}
\end{equation}
where $F_0$ is the baseline fidelity at $z=0$, chosen as the approximate point where the $2$-link curve crosses the $y=x$ threshold in Fig.~\ref{fig:thresholds}. Below this point the encoding no longer provides a net fidelity gain for a $2$-link chain, so it serves as a natural lower bound for the sweep. For idle dephasing, we scale the probability directly as $p_Z^{\mathrm{idle}}(z) = p_{Z,0}^{\mathrm{idle}}(1-z)$ and recover the corresponding $T_2$ from
\begin{equation}
    T_2(z) = \frac{-t_{\mathrm{link}}}{\ln(1 - 2\,p_Z^{\mathrm{idle}}(z))}
    \label{eq:z_t2}
\end{equation}
with $p_{Z,0}^{\mathrm{idle}}$ the baseline dephasing probability at $z=0$, $T_2(0) = 2$~s, and $t_{\mathrm{link}}$ the single-link expected entanglement generation latency. 
$t_{\mathrm{link}}$ here is just for defining the sweep. In the simulation itself, each qubit's idle time is tracked individually based on when things actually happen in the protocol. We cap $T_2$ at $199.99$~s near $z=1$ to fulfill the $2T_1 \geq T_2$ constraint from Eq.~\eqref{eq:idle}, since $T_1 = 100$~s.

\begin{table}[t]
\centering
\caption{Selected hardware parameters in the coordinated $z$ sweep}
\label{tab:z_values}
\fontsize{7}{8.5}\selectfont
\renewcommand{\arraystretch}{1.3}
\begin{tabular}{|c|c|c|c|c|c|c|}
\hline
$\boldsymbol{z}$ & $\boldsymbol{F}_{\mathbf{1q}}$ & $\boldsymbol{F}_{\mathbf{2q}}$ & $\boldsymbol{F}_{\mathbf{m}}$ & $\boldsymbol{F}_{\mathrm{\mathbf{init}}}$ & $\boldsymbol{F}_{\mathrm{\mathbf{phys}}}$ & $\boldsymbol{T}_{\mathbf{2}}$ (s) \\
\hline\hline
0.00 & 0.999000 & 0.999100 & 0.996000 & 0.990000 & 0.965000 & 2.000 \\
\hline
0.25 & 0.999250 & 0.999325 & 0.997000 & 0.992500 & 0.973750 & 2.669 \\
\hline
0.50 & 0.999500 & 0.999550 & 0.998000 & 0.995000 & 0.982500 & 4.008 \\
\hline
0.65 & 0.999650 & 0.999685 & 0.998600 & 0.996500 & 0.987750 & 5.728 \\
\hline
0.80 & 0.999800 & 0.999820 & 0.999200 & 0.998000 & 0.993000 & 10.030 \\
\hline
0.90 & 0.999900 & 0.999910 & 0.999600 & 0.999000 & 0.996500 & 20.068 \\
\hline
0.95 & 0.999950 & 0.999955 & 0.999800 & 0.999500 & 0.998250 & 40.144 \\
\hline
1.00 & 1.000000 & 1.000000 & 1.000000 & 1.000000 & 1.000000 & 199.99 \\
\hline
\end{tabular}
\end{table}

\begin{table}[t]
\centering
\caption{Logical-protocol resource counts per protocol for an $N$-node linear chain, $N\ge2$}
\label{tab:circuit_resources}
\fontsize{7}{8.5}\selectfont
\renewcommand{\arraystretch}{1.4}
\begin{tabular}{|l|c|c|}
\hline
\textbf{Resource} & \textbf{\texttt{QRE-CEC}} & \textbf{\texttt{QRE-None}} \\
\hline\hline
Total Qubits      & $30(N-1)$ & $30(N-1)$ \\
\hline
Total Operations  & $93N-121$ & $93N-121$ \\
\hline
2Q Gates          & $43N-50$  & $43N-50$ \\
\hline
1Q Gates          & $20N-27$  & $20N-27$ \\
\hline
Measurements      & $30N-44$  & $30N-44$ \\
\hline

\end{tabular}

\end{table}

\subsubsection{Resource Requirements}
Each $\llbracket 7,1,3 \rrbracket$ Steane-encoded block requires seven data qubits, seven communication qubits, and one ancilla qubit for FT-Minimal state preparation~\cite{goto2016minimizing}. Both \cec\ and \none\ have the same quantum resource counts because classical error correction only acts on measurement outcomes. 
Table~\ref{tab:circuit_resources} lists these counts as a function of the number of nodes $N$. 
In practice, the runtime is probabilistic since FT-Minimal preparation can require repeated attempts and heralded entanglement generation at the links has variable latency. 
As $N$ grows, the circuit gets deeper, so gate and measurement errors contribute more to the final infidelity. 
The goal of our simulations is to find the regime where error suppression from encoding outweighs the cost of deeper circuits.

\subsubsection{Compared Methods}
We evaluate two protocols. Our \cec, in which qubits are encoded into the $\llbracket 7,1,3 \rrbracket$ Steane code, encoded swapping is performed at the logical level via transversal CNOT gates, and classical error correction is applied to the BSM outcomes. 
Bare \none\ serves as the baseline and includes the same encoding generation and swapping but lacks classical error correction. 
The metrics that are compared are end-to-end logical Bell pair fidelity and latency. Fidelity is the average fidelity of tens of thousands of end-to-end logical Bell pairs generated. 
Latency is also averaged from those end-to-end logical Bell pairs and is measured from the start of a run at the initiator to the completion of the final 2-bit Pauli frame update.

\subsection{Simulation Results}

We first verify that the protocol achieves second-order error suppression across all noise parameters, then study how fidelity and latency depend on network topology, link distance, and overall hardware quality.

\subsubsection{Second-Order Error Suppression Across Simulation Parameters}

Figure~\ref{fig:thresholds} sweeps each of the six physical error parameters in isolation for linear topologies with $1$, $2$, and $5$ links, with all other noise set to zero.

In (a)--(f), the dashed $y = x$ line is the break-even threshold. When a curve sits above this line, the end-to-end logical fidelity exceeds the raw physical parameter being swept, meaning the encoding provides a net fidelity gain at that operating point. All six parameters eventually cross above $y = x$ once the physical fidelity is high enough. Comparing across subfigures, parameters whose curves cross the $y = x$ line further to the right correspond to noise sources that the protocol is more resilient to, since the encoding breaks even at higher physical error rates.
The log--log plots (g)--(l) make the suppression of error scaling explicit. The protocol in~\cite{jiang2007distributed} claims that for an $\llbracket n, k, 2t+1 \rrbracket$ CSS code with classical error correction, the logical error rate is suppressed to $Q \sim q^{t+1}$ where $q$ is the effective error probability per physical qubit. For the Steane $\llbracket 7,1,3 \rrbracket$ code with $t = 1$ this means $Q \sim q^2$. Our simulation confirms this prediction. On the log--log plots, the fitted slope $c$ measures the exponent in $O(\epsilon^c)$ scaling of the logical error rate with physical error rate $\epsilon$. Without encoding, errors accumulate as $O(\epsilon)$. The fitted slopes in the legends range from $1.81$ to $2.19$ across all sweeps and topologies, consistent with $O(\epsilon^2)$ and confirming approximate second-order suppression for every noise source we tested.

\begin{figure}[h]
    \centering
    \includegraphics[width=\linewidth]{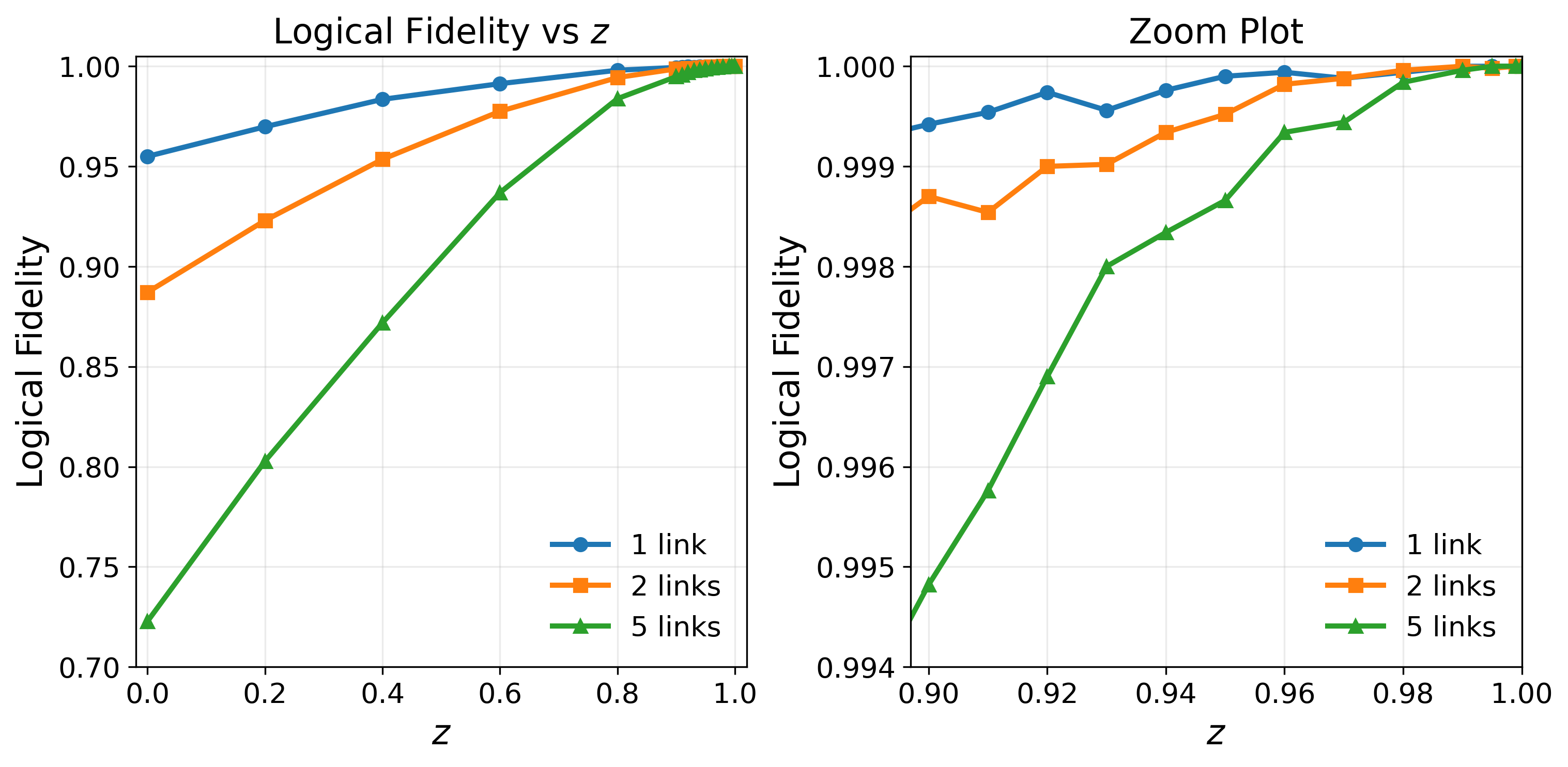}
    \vspace{-0.3in}
    \caption{End-to-End Logical Bell-pair Fidelity versus the hardware sweep parameter $z$ for $1$-, $2$-, and $5$-link linear repeater chains with fixed 20~km inter-node spacing. Higher $z$ corresponds to uniformly better hardware performance across all swept parameters. The right panel zooms into the near-perfect-fidelity regime.}
    \vspace{-0.2in}
    \label{fig:z_sweep}
    % \vspace{-0.2in}
\end{figure}

\subsubsection{Combined Hardware Sweep}

Figure~\ref{fig:z_sweep} sweeps all hardware noise together using the $z$ parameter from Eq.~\eqref{eq:z_fidelity}--\eqref{eq:z_t2} and Table~\ref{tab:z_values}, for $1$-, $2$-, and $5$-link chains. 

At $z=0$ the $1$-link chain is at $0.953$ while the $5$-link chain is at $0.72$, a gap of $0.23$ driven entirely by the extra circuit depth in longer chains. By $z=0.8$ this gap has narrowed to about $0.01$, meaning that better hardware disproportionately helps the longer chains that need it most. Past $z=0.96$ all three topologies are above $0.999$ and the curves flatten out, so there are diminishing returns from pushing hardware further once you are in this regime. The noisiness of the zoom plot (right) is from the fidelities being so close to $1$ that more Monte Carlo samples would be needed to see smooth curves. The practical implication is that chain length matters a lot at mediocre hardware quality but becomes almost irrelevant once the hardware is good enough.

\subsubsection{Choosing the Best Topology for a Given Distance}

\begin{figure}[h]
    \vspace{-0.1in}
    \centering
    \includegraphics[width=\linewidth]{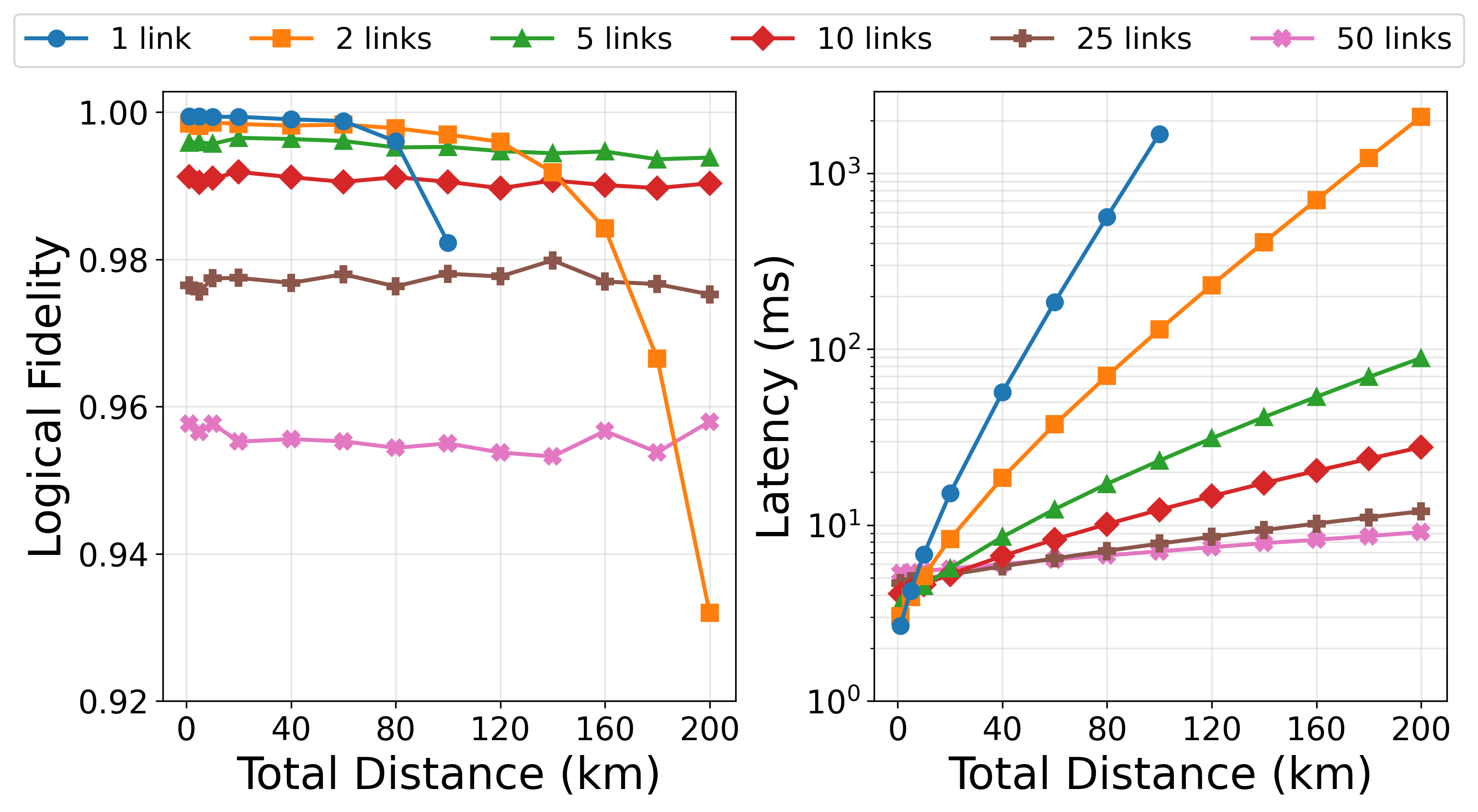}
    \vspace{-0.2in}
    \caption{Logical fidelity (left) and latency (right) versus total end-to-end distance for $1$-, $2$-, $5$-, $10$-, $25$-, and $50$-link repeater chains at the shared $z=0.9$ operating point. The latency axis is shown on a logarithmic scale.}
    \label{fig:distance_topology_sweep}
\end{figure}

Figure~\ref{fig:distance_topology_sweep} shows logical fidelity and latency versus total distance for $1$- through $50$-link chains, all at $z=0.9$.

No single topology is optimal everywhere. The $1$-link chain stays above $0.99$ out to $80$~km, but by $100$~km the latency reaches $\sim 1.5$~s and fidelity drops as qubits idle for a noticeable fraction of $T_2 \approx 20$~s. The $2$-link chain pushes this cliff out to about $160$~km before the same degradation is experienced.
The $10$- and $50$-link chains have low latency even at long distances because each elementary link is short, but their fidelities are lower across the board due to the larger number of gates and measurements. This is consistent with the $z$ sweep (Fig.~\ref{fig:z_sweep}), where longer chains needed higher $z$ to reach the same fidelity.

The choice of topology depends on what dominates at a given target distance, idling decoherence from few long links or accumulated gate errors from many short ones. Below $80$~km, $1$ or $2$ links gives the best fidelity. Between $140$ and $200$~km, $5$ to $10$ links is the better trade-off. Beyond that, both more links and better hardware are needed.

\subsubsection{Effect of Network Scale}

\begin{figure}[h]
    \centering
    \vspace{-0.1in}
    \includegraphics[width=\linewidth]{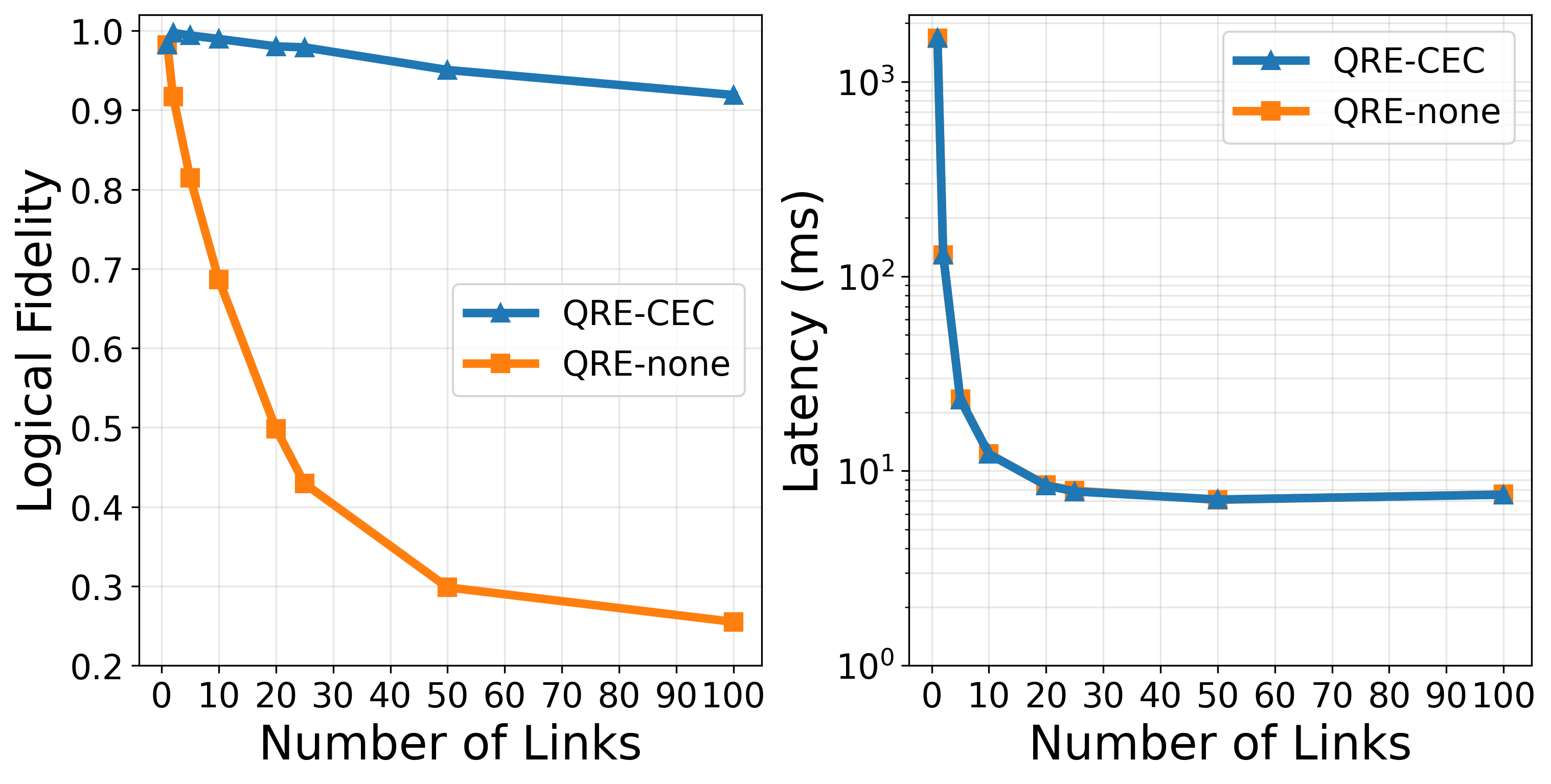}
    \vspace{-0.25in}
    \caption{End-to-end fidelity (left) and latency (right) versus number of links
    over a fixed 100~km total distance.}
    \label{fig:num_links}
    
    % \vspace{0.1in}
\end{figure}

\begin{figure}[ht]
    \centering
    \includegraphics[width=\linewidth]{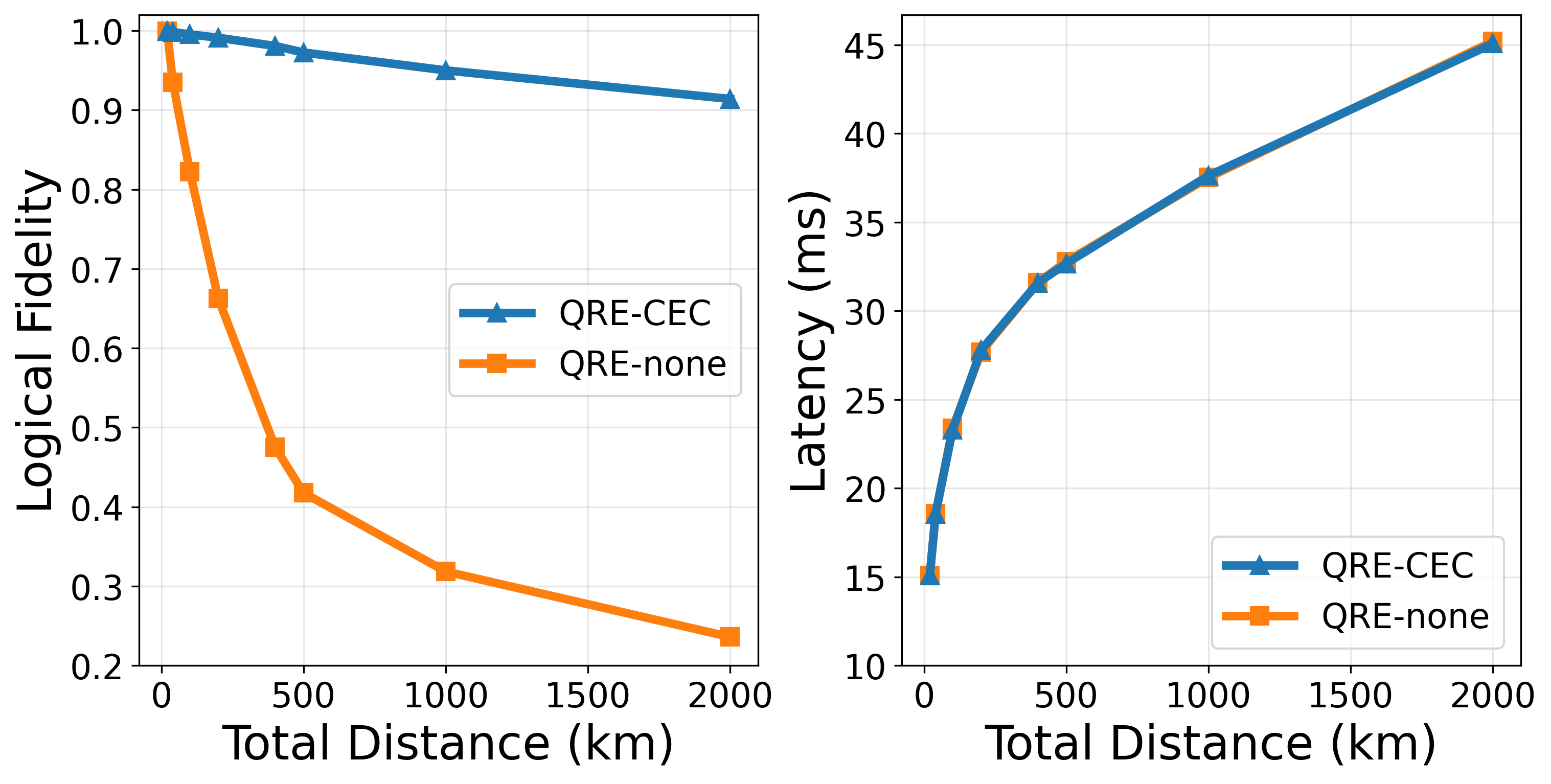}
    \vspace{-0.25in}
    \caption{End-to-end fidelity (left) and latency (right) versus total distance with fixed 20~km inter-node spacing.}
    \label{fig:distance}
    % \vspace{-0.2in}
\end{figure}

Figures~\ref{fig:num_links} and~\ref{fig:distance} compare \cec\ and \none\ as network size grows.

Figure~\ref{fig:num_links} holds the total distance at $100$~km and increases the number of links. At $1$ link, both protocols give $\sim 0.98$ fidelity because there are no intermediate nodes for correction to act on. 
Once intermediate nodes appear, there is a gap in performance between the two. As links increase, \none\ drops rapidly and reaches the fully mixed value of $0.25$ at $100$ links, while \cec\ only falls to $0.91$, indicating that classical error correction is catching most single-qubit errors at each intermediate node. Latency is identical for both protocols since classical correction overhead is negligible.

Figure~\ref{fig:distance} holds the inter-node spacing at $20$~km and increases total distance. Fidelity for \cec\ is above $0.99$ out to $100$~km (5 links) and still at $0.91$ at $2000$~km (100 links). 
\none\ falls to $0.83$ at $100$~km and $0.24$ at $2000$~km. That \cec\ remains above $0.9$ over $100$ hops is consistent with second-order error suppression from the Steane code preventing per-hop errors from compounding. 
Latency grows linearly from $\sim 15$~ms to $\sim 45$~ms for both protocols. The gradual decline of \cec\ fidelity toward $2000$~km reflects the accumulation of uncorrectable weight-two errors over a large number of hops, suggesting that longer distances could benefit from codes with larger code distances ($d \ge 3$) that further suppress error per hop.

\section{Conclusion and Future Work}
\label{sec:conclusion}
Our simulation of the \cec\ protocol in SeQUeNCe shows that combining quantum encoding with classical error correction can produce high-fidelity logical Bell pairs at low latency over long distances in the parameter regimes we studied. Across all parameter sweeps, \cec\ consistently achieves higher end-to-end logical Bell-pair fidelity than \texttt{QRE-None}.

More broadly, the performance of encoded quantum repeater networks depends on the interplay between circuit depth and latency. Deeper circuits from more intermediate nodes accumulate more gate and measurement errors, while longer latencies from farther-apart repeaters accumulate more idling decoherence. Since photon loss grows exponentially with link distance, spacing repeaters further apart demands higher $T_1$ and $T_2$ to tolerate the longer wait times. Adding more intermediate nodes within a fixed distance can cut latency but increases circuit depth. 
Noisier gates mean more errors to correct, and at some point the correction overhead itself becomes the problem. Balancing circuit depth against latency for a given hardware set is the central design problem for encoded repeater chains. For near-term deployments, gate fidelity is probably the most important parameter to improve since coherence times stop being the bottleneck well before gate noise does.

Seven future directions are worth pursuing.
(1) Lighter codes such as the 3-qubit or 5-qubit repetition codes may outperform the Steane code in high-noise regimes despite weaker correction, since they need fewer gates. (2) Codes with larger distance, since error suppression scales as $q^{t+1}$ for an $\llbracket n, k, 2t+1 \rrbracket$ CSS code and larger $t$ would push the logical error rate down faster.
(3) To keep the logical Bell pair generation rate constant as distance grows, as in Figure 1 of~\cite{jiangencoding-pra-2009}, the end repeaters need enough quantum memories to buffer pairs that have not yet received their Pauli frame corrections. Managing this buffering is an open problem in protocol design.
(4) The hybrid strategy in~\cite{hybrid-qcnc2024} that combines entanglement distillation from 1G repeaters with error correction from 2G repeaters could be integrated into \cec.
(5) Reducing the idling decoherence can be achieved by using more communication qubits, such that extra physical Bell pairs are generated. This can reduce the expected latency to generate the needed amount of physical Bell pairs.
(6) Replacing the depolarizing gate noise with the biased Pauli channel already implemented in the simulator would give a more realistic picture, since real hardware typically has a strong $Z$ bias from dephasing. Asymmetric codes designed to exploit this bias could then outperform the symmetric Steane code by dedicating more correction power to the dominant error type.
(7) A natural application is quantum key distribution~\cite{bennett-qkd-14}, where the fidelity and latency numbers reported here translate directly to secret key rate and communication distance.

With continued hardware improvements and smart protocol-level choices like code selection, fault-tolerant preparation, and hybrid correction strategies, encoded repeater networks can realistically deliver high-fidelity entanglement over continental distances.

\bibliographystyle{IEEEtran}
\bibliography{bib}

\end{document}